\documentclass[11pt,a4paper]{article}
\usepackage{caltn,psfig}
\usepackage{graphicx}
\usepackage{hyperref}

\begin{document}

\docnum{IACHEC Report 2018}
\title{Summary of the 13th IACHEC Meeting}
\author{K. K. Madsen$^a$, L. Natalucci$^b$, G. Belanger$^c$, C. E. Grant$^d$, M. Guainazzi$^e$, V. Kashyap$^f$,  \\
 H. L. Marshall$^d$, E. D. Miller$^d$, J. Nevalainen$^g$, P. P. Plucinsky$^f$, Y. Terada$^h$ \\ \\
  $^a$Cahill Center for Astronomy and Astrophysics, California Institute of Technology, USA \\
  $^b$IAPS-INAF, Italy \\
  $^c$ESA-ESAC, Spain \\
  $^d$Kavli Institute for Astrophysics and Space Research, Massachusetts Institute of Technology, USA \\
  $^e$ESA-ESTEC, The Netherlands \\
  $^f$Harvard-Smithsonian Center for Astrophysics, USA \\
  $^g$University of Tartu, Estonia \\
  $^h$University of Saitama, Japan
}
\date{\today}                   

\newcommand{\xmm}{XMM-\textit{Newton}}
\newcommand{\chandra}{\textit{Chandra}}
\newcommand{\suzaku}{\textit{Suzaku}}
\newcommand{\swift}{\textit{Swift}}
\newcommand{\nicer}{\textit{NICER}}
\newcommand{\astrosat}{\textit{Astrosat}}
\newcommand{\nustar}{\textit{NuSTAR}}
\newcommand{\hxmt}{\textit{Insight-HXMT}}
\newcommand{\hitomi}{\textit{Hitomi}}
\newcommand{\integral}{\textit{INTEGRAL}}
\newcommand{\fermi}{\textit{Fermi}}
\DeclareRobustCommand{\ion}[2]{\textup{#1\,\textsc{\lowercase{#2}}}}
\newcommand{\cstat}{{\tt c-stat}}

\maketitle

{\small
\centering
{\bf Abstract} \\
We summarize the outcome of the 13th meeting of the International Astronomical Consortium for High Energy Calibration (IACHEC), held at \textit{Tenuta dei Ciclamini} (Avigliano Umbro, Italy) in April 2018. Fifty-one scientists directly involved in the calibration of operational and future high-energy missions gathered during 3.5 days to discuss the current status of the X-ray payload inter-calibration and possible approaches to improve it. This summary consists of reports from the various working groups with topics ranging from the identification and characterization of standard calibration sources, multi-observatory cross-calibration campaigns, appropriate and new statistical techniques, calibration of instruments and characterization of background, and communication and preservation of knowledge and results for the benefit of the astronomical community.
}

\section{Introduction}

The International Astronomical Consortium for High Energy Calibration (IACHEC)\footnote{\href{http://web.mit.edu/iachec/}{\tt http://web.mit.edu/iachec/}} is a group dedicated to supporting the cross-calibration of the scientific payload of high energy astrophysics missions with the ultimate goal of maximizing their scientific return. Its members are drawn from instrument teams, international and national space agencies, and other scientists with an interest in calibration. Representatives of over a dozen current and future missions regularly contribute to the IACHEC activities. Support for the IACHEC in the form of travel costs for the participating members is generously provided by the relevant funding agencies.

IACHEC members cooperate within Working Groups (WGs) to define calibration standards and procedures. The objective of these groups is primarily a practical one: a set of data and results produced from a coordinated and standardized analysis of high-energy reference sources that are in the end published in refereed journals. Past, present, and future high-energy missions can use these results as a calibration reference.

The 13th IACHEC meeting was hosted by Lorenzo Natalucci (IAPS-INAF), and was held at  \textit{Tenuta dei Ciclamini} (Avigliano Umbro) in Italy. The meeting was attended by 51 scientists from the US, Europe, and Asia representing high energy missions based in US, UK, Italy, Germany, France, Netherlands, Switzerland, Spain, China, India, and Japan. The meeting also featured presentations from IACHEC members unable to attend in person who were able to view and contribute to the meeting via video teleconferencing. Advances in the understanding of the calibration of more than a dozen missions was discussed, covering multiple stages of operation and development:

\begin{itemize}
  
\item[-] Recently completed missions - \suzaku\ and \hitomi
\item[-] Currently operating missions - \chandra, \xmm, \swift, \integral, \astrosat, \nustar, \nicer, and \hxmt\
\item[-] Pre-launch status - eROSITA
\item[-] Missions under development: \textit{IXPE}, \textit{XRISM}, and {\it Athena}
\item[-] Ground calibration facilities: acceleration facilities at the IAAT of the University of T\"ubingen, GSFC/CREEST, MPE/Panter, NASA/XRCF among others
\item[-] Atomic databases: AtomDB

\end{itemize}

This year, the post of IACHEC chair was passed from M. Guainazzi to K. Madsen after 10 years of service.

This report summarizes the main results of the 13th meeting and comprises the reports from each of the IACHEC WGs. The presentations made at the meeting are available at: \\ {\href{http://web.mit.edu/iachec/meetings/2018/index.html}{\tt http://web.mit.edu/iachec/meetings/2018/index.html}}.

The IACHEC gratefully acknowledges sponsorship for the meeting from ESA ESTEC/Science Support Office and from the European Union's Horizon 2020 Programme under the AHEAD project (Grant agreement n. 654215). 

\section{Working Group reports}

\subsection{Calibration Uncertainties: CalStats}\label{s:calstat}

The CalStats Working Group was designed as a forum to discuss statistical, methodological, and algorithmic issues that affect the calibration of astronomical instruments, as well as how they are used in data analysis and the interpretation of analysis results. During this workshop, four events were of direct relevance: two working group sessions and two talks during plenary sessions. The topics ranged from discussions of the Concordance Project (\S\ref{s:concordance}) to practical binning of spectra (\S\ref{s:spectra}) and use of the \cstat\ statistic in spectral fitting (\S\ref{s:cstat}), and included a tutorial on Gaussian Processes (\S\ref{s:GP}).

\subsubsection{Concordance}\label{s:concordance}

The Concordance Project is an effort to address one of the primary objectives of IACHEC, which is to determine the magnitude of the systematic corrections to the sensivities of the various instruments. Several sets of observations of the same sources have been carried out with multiple instruments, either simultaneously or contemporaneously for those with low intrinsic variability, that it is feasible to compare the predicted fluxes and determine the magnitude of deviations from each other (see, e.g., the comprehensive modeling of the supernova remnant 1E 0102.2-7219 in the Small Magellanic Cloud, by Plucinsky et al.\ 2017). The question that arises, however, is to determine the expected base level of predicted fluxes when all the measurements have different statistical and expected systematic errors. The strategy that has been used to determine this base level is mathematical shrinkage, whence multiple estimates are combined with different weights of statistical and systematic errors, with the full dataset driving the precise weight that determines how much of the prior information is combined with the values directly estimated from the data.

The analysis requires both observations of common sources as well as prior estimates from the instrument team as to the magnitude of anticipated systematic error (colloquially called $\tau$'s in honor of arcane notation) in each instrument. The final estimate gives an indication of how much of the estimated correction can be attributed to the data and how much to the anticipated systematics. This allows individual calibration teams to judge whether any changes to their effective area estimates are required, since large estimated corrections that are attributable to prior systematics are unlikely to be convincing.

Over the past two years, the Concordance team has been collecting $\tau$'s for instruments and passbands of interest to IACHEC. More were collected after Herman Marshall's progress report (Session {\tt VII}/Special Session on the Concordance project). A paper describing the mathematical foundation of the method has been accepted for publication in a statistics journal (Chen et al.\ 2018). An additional paper that applies the method to a larger variety of datasets and describes the results from an astronomer's viewpoint is in preparation (Marshall et al.\ in prep).

\subsubsection{Spectral Binning}\label{s:spectra}

The new generation of X-ray telescopes that were spearheaded by {\sl Hitomi} ({\sl Arcus}, {\sl XRISM}, {\sl Athena} {\sl Lynx}) illustrate the importance of high-resolution spectral analysis. Unlike with the gratings on \chandra\ and \xmm, this new generation of telescopes will allow observations of significantly weaker sources, expanding the space of high-quality data by orders of magnitude. Analyzing these data remains a challenge from several perspectives: first, these data will require a vast improvement in the quality of the astrophysical emission models that are used to fit to them; second, analysis algorithms that can take advantage of the high resolution and the availability of increased information will have to be developed; and third, analysis environments like {\sl XSPEC} and {\sl CIAO/Sherpa} must adapt to the larger data sizes. 

A way to improve the handling of the last of these issues is the focus of Kaastra \& Bleeker (2016). They have devised a method to improve upon how much of a Redistribution Matrix File (RMF) needs to be stored, and how much its use can be improved by accounting for how spectral responses behave across the finite widths of the bins in the stored file. They describe this as a method to determine the optimal binning of model spectra, and find that when higher order corrections to the response within a bin are included, larger bin widths are adequate to describe the RMF.

These tactics are currently available in {\sl SPEX} and are being implemented in {\sl XSPEC}. Detailed prescriptions for how to determine the optimal binning are given in the Appendix of Kaastra \& Bleeker (2016). J. Kaastra gave a plenary talk (Session {\tt IX}/Special Session on IXPE, XARM, and X-ray spectroscopy statistics) on this subject and also continued discussion during the second CalStats Working Group session (Session {\tt XI}).

\subsubsection{Fitting via \cstat}\label{s:cstat}

Astronomers have historically tended to use the $\chi^2$ statistic as a fitting metric as well as a measure of the goodness of the fit. However, it is not an optimal measure for the types of data encountered in high-energy astronomy. When there are many bins with few to no counts, the Poisson nature of the data cannot be ignored and it is advantageous to use the \cstat\ statistic, which is a form of the Poisson log-Likelihood. This produces unbiased parameter estimates, even in the low counts limit, without having to group or bin up the data at the cost of information present at higher resolutions. Unfortunately, while it is asymptotically distributed as a chi-square statistic, in the regular case a goodness of fit cannot be inferred based on the value of the fit statistic. Kaastra (2017) set out to remedy this, and derived the {\sl expected} value of the \cstat\ and its variance in a given bin for a large range of model intensity values. Then, for a given spectral array, it is a simple matter to add up the expected values and variances for the best-fit model, and compare it to the actual value of the \cstat. How far the observed \cstat\ lies from the expected value, in terms of the expected standard deviations, measures how good the model fit is.
Kaastra (2017) gives a comprehensive piecewise-functional description of the expected values and variances, which can be easily implemented.\footnote{See, e.g., \href{https://github.com/vkashyap/PINTofALE/blob/master/pro/stat/cstat\_gof\_k17.pro}{\tt https://github.com/vkashyap/PINTofALE/blob/master/pro/stat/cstat\_gof\_k17.pro} from PINTofALE.} This is a promising development, and was discussed extensively during the second CalStats Working Group session (Session {\tt XI}). Efforts are underway by several statisticians to develop a more principled justification for the formulation.

\subsubsection{Gaussian Processes}\label{s:GP}

The first working group session (Session {\tt III}/CalStats) was devoted to a tutorial on Gaussian Processes, conducted remotely by statistician David Jones (SAMSI/Duke \& TAMU).

New methodologies to do astronomical analysis are coming up rapidly. The most prominent of these are Machine Learning methods, which include Neural Networks (also known as Deep Learning) and Gaussian Processes (GPs). Unlike traditional statistical analysis, they seek to derive an accurate phenomenological description of the data that can be used to understand how a system might evolve despite having an incomplete physical understanding of the system. They are fully data driven, and provide a useful check against model-oriented analyses, and are thus most immediately useful to address calibration problems. For example, GPs (whose mentions in astronomical literature has been growing exponentially in recent years) are useful to obtain smoothed semi-parametric representations of the data with relatively high predictive power. GPs essentially model a response variable as a multinomial Normal distribution, allowing us to predict the values of a smoothly varying function at points where no observations have been carried out. The slides from the tutorial are available at the IACHEC website\footnote{(\href{http://web.mit.edu/iachec/meetings/2018/presentations/Jones_SessionIII.pdf}{\tt http://web.mit.edu/iachec/meetings/2018/presentations/Jones\_SessionIII.pdf})} and include several references to additional resources (e.g., Rasmussen \& Williams, 2006)

\subsection{Contamination}
The Contamination Working Group met for one session of updates from recently operating missions.  As at previous IACHEC workshops, the group discussed affects of molecular contamination on soft X-ray instruments (e.g., Marshall et al. 2004, Koyama et al. 2007, O'Dell et al. 2013).  Since its inception, the Working Group has covered three broad topics: (1) comparison of contamination among instruments and missions; (2) mitigation for current instruments; and (3) mitigation for future instruments.

Since \chandra\ is the only operating X-ray observatory suffering greatly from molecular contamination, the entire 2018 WG session was devoted to updates on \chandra\ ACIS contamination monitoring and characterization, much of which is detailed by Plucinsky et al. (2018).  A.\  Bogdan presented recent raster-scan observations of the galaxy cluster Abell 1795, a standard contamination calibration target, and results from the field-filling External Calibration Source (ECS).  Both datasets suggest that, while contamination appears to still be accumulating on the ACIS optical blocking filter (OBF), the rate of contamination may be slowing down.  The centers of the ACIS-I and ACIS-S array filters are behaving differently.  The accumulation rate at the center of the ACIS-I array filter from 2017--2018 is significantly less than what was in the N0010 contamination model released by the CXC and significantly less than a linear extrapolation of the earlier data.  In fact, the 2017--2018 data are consistent with little or no accumulation.  The accumulation rate at the center of the ACIS-S array filter from 2017--2018 is consistent with a linear extrapolation of the earlier data but still significantly less than what was in the N0010 contamination model. The newly accumulating contaminant appears less sensitive to temperature differences between the center and edge of the filter.  Additionally, the contamination has reached a level where ECS Al K lines at 1.5 keV can now be used to track the spatial structure, using the ratio of Al K to the Mn K (5.9 keV) line strength, as the latter is unaffected by contamination.  

H.\ Marshall added updates from regular ``big dither'' observations of blazars using LETG/ACIS.  These observations involve a large dither of the bright continuum source Mkn 421 along the Y (cross-dispersion) direction on ACIS-S to efficiently measure the depth of the contaminant C, O, and F K edges as a function of distance from the filter edge.  Among other results, the big dither suggests there are two components of the contaminant with different growth times: a spatially non-uniform component with a long growth time that has possibly leveled off, and a uniform component with a short growth time that is currently building up.

Finally, P.\ Plucinsky presented new observations of 1E~0102.2-7219, an SNR target used to verify the contamination model due to its strong, isolated He-like and H-like O and Ne lines.  In these observations from 2017--2018, ACIS-I3 shows an increase in the line normalizations compared to the prediction of the extrapolated contamination model, again indicating that the accumulation rate was overestimated in the N0010 contamination model.  ACIS-S3 shows a similar trend but not as large in the center of the CCD, with the bottom of the CCD consistent with previous line normalization measurements. Since the IACHEC meeting, the ACIS contamination model has been updated twice, to N0012, including improvements to both the ACIS-I and ACIS-S models which reduce the overestimatation.  CALDB version 4.8.1 contains both these updates.

The WG plans to continue sharing and discussing contamination monitoring on existing missions.  In addition, since much of the recent contamination analysis is limited to unrefereed proceedings, technical memos, and user documentation, calibration scientists from several missions indicated their plans to submit manuscripts to refereed journals over the course of the next year.

\subsection{Coordinated Observations}
The objectives of the IACHEC Coordinated Observations Working Group are to coordinate new observations jointly among different telescopes, analyze those observations, and publish the results. For the 2018 IACHEC meeting, there were two presentations on cross-calibration projects and significant discussion of potential new observations.

K.\ Madsen reported results from work on joint \nustar\ and \swift\ observations of bright Galactic X-ray binaries.  The largest differences were at the low energy end, which were previously interpreted as resulting from dust haloes for sources with very high neutral column density.  Examples were GX 13+1 and GRS 1915+105.  Independent reductions from different people do in some cases give different results, and it seems that some of the differences between the two observatories may be due to analysis parameters such as the chosen extraction region and the excision of the central core in \swift, which in the case of dust scattering haloes result in spectra that can not be directly compared.  One conclusion from this work is that we need to establish and publish uniform analysis procedures.

L.\ Natalucci provided an update to the analysis of joint \nustar\ and \integral\ observations of 3C~273 from campaigns in 2012, 2015, 2016, and 2017.  In particular, he concentrated on the flux in the 20--40 keV band, finding that the two telescopes were consistent at the 10\% level, consistent with the uncertainties.  Spectral slopes were generally consistent.

The working group established a simple procedure for handling coordinated observations and the resultant data.  A person will be identified who will lead the analysis, collecting good time intervals (GTIs) and computing the overlaps between observatories, as was done in the campaigns published by Madsen et al. (2017).  The GTI overlaps would then be provided to the contacts for each project involved and these contacts will extract spectra and generate response functions, which will then be passed to the coordination lead for bandpass analysis.  Analysis leads were established: K.\ Madsen for 2015, 2016, and 2017 observations of 3C~273, C.\ Markwardt for the 2018 observation of 3C~273, N. Schulz for GX~13+1, E. Jourdain for MAXI J1820, V.\ Kashyap and J.\ Kaastra for many \chandra\ and \xmm\ observations of Capella, and P.\ Kretschmar for Her~X-1.  The contacts for various instruments are: H.\ Marshall for \chandra, E.\ Jourdain for \integral/SPI, L.\ Natalucci for \integral/IBIS, M.\ Stuhlinger for \xmm, A.\ Beardmore and J.\ Kennea for \swift, K.\ Madsen and K.\ Forster for \nustar, C.\ Markwardt for \nicer, and L.\ Ming Song for \hxmt.

\subsection{Detectors and Background}
The Detectors and Background Working Group met in one well-attended session. As always, the Detectors Working Group provides a forum for cross-mission discussion and comparison of detector-specific modeling and calibration issues, while the Background Working Group provides the same for measuring and modeling instrument backgrounds in the spatial, spectral, and temporal dimensions. Attendees represented many X-ray missions, including \xmm, \chandra, \nustar, and \hxmt.

The first session began with an introduction by Y.\ Jiang to the Low Energy detector on China's \hxmt, which makes use of Swept Charge Devices. He described the ground calibration effort to measure energy scale and FWHM and their dependence on detector temperature, as well as QE and detector background. \hxmt\ was launched in Summer 2017 and early on-orbit calibration results were also presented.

Next were two presentations describing on-going \chandra\ ACIS calibration activities. T.\ Gaetz spoke on the status of updating the response and gain, in particular correcting a droop in the energy scale calibration for a narrow region near the central node boundary of the front-illuminated CCDs. N.\ Durham discussed improvements and updates to the quantum efficiency uniformity maps. Both updates will be available in a future release of the \chandra\ CALDB.

K.D.\ Kuntz described what is known about the sources of the \xmm\ particle background and in particular the quiescent particle background which is primarily due to Galactic Cosmic Rays. He examined evidence for variability in the spectrum and the hardness ratio of the quiescent background which is not clear at this point. Residuals due to other components, such as trapped radiation and soft proton flares, are difficult to completely remove.

B.\ Grefenstette reviewed \nustar\ CZT gain monitoring. The initial gain was determined during ground calibration and has been verified twice by deploying an Eu-155 source on-orbit. In addition, time-dependence of gain has been calibrated using background lines at three energies to monitor the slope, and observations of the Kepler SNR to monitor the offset. The current CALDB reproduces the high-energy background line energies very well, with continued monitoring to explore detector-to-detector variations  and the origin of long-term variation.

Finally, C.\ Heinitz reported on CORRAREA, an empirical correction of the EPIC on-axis effective areas with the goal of producing consistent effective areas for both EPIC-MOSs and EPIC-pn. Sources are selected from the \xmm\ Serendipitous Source Catalogue and then stacked for each detector. A phenomenological reference model is fit to the EPIC-pn data, convolved with the instrument responses, then the residual ratio is calculated and fit. Future plans are for more automation and validation. CORRAREA is currently available as a non-default option in the \xmm\ Science Analysis System (SAS).

\subsection{Galaxy Clusters}
The WG mainly discussed the Multi-Mission Study (MMS), a project whereby we aim at comparing X-ray spectroscopic results of a sample of clusters
obtained with five on-going and past X-ray missions and nine instruments.

J.\ Nevalainen presented preliminary results on the comparison of ROSAT/PSPC and \xmm/EPIC-pn instruments using a subsample of 12 clusters. The study of
the shapes of the effective area indicated no energy-dependent cross-calibration effects between the two instruments in the overlapping 0.5--2.0 keV band. However, the pn yielded systematically and significantly a $\sim10\%$ flux excess, compared to the PSPC. If the PSPC calibration is absolutely calibrated, the pn effective area normalization in the soft band would be about 10\% too low. This is different from the previous WG work whereby the comparison with \xmm/MOS and \chandra/ACIS instruments indicated rather a 5\% too high normalization for the pn instrument (Nevalainen et al., 2010) in the soft X-ray band. Also, in the above work the shape of the soft band effective area
calibration was significantly inconsistent between pn and ACIS. More work is required to understand the big picture.

The WG pointed out a possible problem on the scaling of the spectra obtained with different instruments. In particular in the \xmm/EPIC instruments (the pn and the MOS) a significant fraction of the FOV is obscured by the CCD gaps. At the moment we scale the flux linearly with the obscured area fraction to obtain the flux in the full unobscured extraction region. This is accurate only for spatially constant emission, whereas the cluster flux decreases with distance from the cluster center. To proceed on this front, J. Nevalainen agreed with ESA Instrument and Calibration Scientist I.\ Valtchanov to a 2 week visit at ESAC to evaluate the problem in EPIC and, if necessary, to develop a SAS-integrated tool which considers the brightness distribution of a given cluster to fill the CCD gaps with a model-based contribution.

\subsection{Heritage}
 The Heritage Working Group has the scope of preserving the IACHEC corpus of knowledge, know-how, and best-practices for the benefit of future missions and the community at large, by:
\begin{itemize}
\item providing a platform for the discussion of experiences coming from operational missions,
\item facilitating the usage of good practices for the management of pre- and
  post-flight calibration data and procedures, and the maintainance and
  propagation of systematic uncertainties (the latter task in strict
  collaboration with the Systematic uncertainties IACHEC Working Group),
\item documenting the best practices in analyzing high-energy astronomical data as a
  reference for the whole scientific community,
\item ensuring the usage of homogeneous data analysis procedures across the IACHEC
  calibration and cross-calibration activities,
\item consolidating and disseminate the experience of operational missions on the
  optimal calibration sources for each specific calibration goal.
\end{itemize}

The main current activity of the WG is the development of the IACHEC Source Database (ISD). The ISD is defined as the single repository of high-level scientific data and data analysis procedures used in IACHEC published papers. The ISD shall be populated by an IACHEC Working Group whenever:

\begin{itemize}
\item An IACHEC paper is published on a refereed journal,
\item Updated calibrations with respect to that used in the aforementioned paper
  are published, and they have a significant impact on the results of the
  cross-calibration analysis, as verified by the Working Group and documented
  in a Technical Note, or in a new paper.
\end{itemize}

Papers and Technical Notes are supposed to be also ingested in the ISD, besides being available from the IACHEC web portal.

A first version of the database interface was presented at the 12th IACHEC meeting, and discussed with scientists from all IACHEC Working Groups. It is expected that the ISD will be open to the community in the coming months.

The implementation of the ISD was made possible thanks generous support by the AHEAD Project funded by the European Union.

\subsection{Non-Thermal SNRs}
The meeting of the Non-Thermal SNR WG has been mostly devoted to results of the Crab obsevations. L. Natalucci presented the progress of the the ``multi-year" cross-calibration project, taking advantage of a sample of 14 nearly simultaneous observations during a decade spanning 2005--2016. The sample currently includes data from \nustar, \integral/IBIS and SPI, RXTE/PCA, \suzaku/XIS and HXD, \swift/BAT and \fermi/GBM. The spectral fits with a power law (or broken power law model, for energy coverage beyond 100~keV) give consistent results for all observations, considering the assumptions in the calibration of the different instruments.

L. Kuiper presented new results on the Crab absolute timing, based on the analysis of the Crab light curve from different high energy instruments. The timing signals from the X-ray and gamma-ray instruments are found to be ahead of the radio signal by $\sim0.2-0.4$~$\mu$s. Within this range, the distribution of this time difference for a given instrument differs significanly from other instrument's measurements, with a possible intrinsic energy dependence.

The \nustar\ team results on the recent Crab campaign were presented by K. Madsen. A dataset of 24 observations spanning two years from April 2016 has been used with the Crab both in nominal pointing and straylight (SL) positions. This has been used to provide complementary monitoring with \fermi/GBM and \swift/BAT and also to calibrate the instrument vignetting functions and detector absorption parameters. The new measurements are expected to result in a new issue of CALDB towards the end of the year with a possible re-calibration of the \nustar\ effective area.

A long term analysis of \swift/BAT data was presented by C.\ Markwardt. The main aim is to study the variation and evolution of the gain of the CZT detectors against different parameters including altitude, orbit evolution, etc. The Crab data were used to monitor the response through the mission lifetime, by studying the flux and slope of the spectrum and showed no real evidence of response changes over time. The effect of damage and annealing has also been investigated.

Finally, two non-thermal SNRs (Crab and G21.5) have also been used to calibrate the \hitomi\ instruments. A presentation provided by M. Tsujimoto presented the latest results of the analysis of the \hitomi\ data, also including a complex analysis of the weight of the different calibration uncertainties on the flux normalization.

\subsection{Thermal SNRs}
The thermal SNRs working group met for one session during the 2018 IACHEC  meeting. 
The main topic of discussion was modifications to the standard IACHEC model for the Large Magellanic Cloud supernova remnant N132D. There is a strong desire in the group to generate a model that works well across the 0.3 to 10.0 keV bandpass so that the model has the greatest applicability for current and future missions. Previously the group had focused on generating a model that would work well in the 1.5--4.0 keV bandpass, in particular for the Si, S, and Ar lines.  

E.\ Miller and H.\ Yamaguchi discussed what was learned in the analysis of the \hitomi\ data (\hitomi\ Collaboration et al., 2018) and the \suzaku\ data (Bamba et al., 2018)
They suggest that at least one more line in the Fe K region and a high temperature continuum component need to be added to the IACHEC model. E.\ Miller commented on the complications with the background models for the \suzaku/XIS in the 5.0-10.0 keV bandpass and suggested that more work could be done on these models to improve the characterization of the high temperature component.  It was suggested that someone from the \nustar\ team should join this effort to assist in the characterization of the high temperature component.  After the meeting, B.\ Grefenstette graciously agreed to join this project and to assume responsibility for the analysis of the \nustar\ data. M.\ Freyberg commented on the background and out-of-time events for the EPIC-pn in small window mode.  He would like to propose a dedicated calibration observation with N132D outside the small window mode but on the active region of the pn in the readout direction to understand this background better.  J.\ Kaastra discussed the work he had done with fitting a spline model to a recent EPIC-RGS spectrum of N132D to develop an empirical model. S.\ Chandra volunteered to take the SPEX output file and convert it to something that could be used in XSPEC.  S.\ Sembay discussed the work that he had done creating SIMPUT files that could be used by the SIXTE software to simulate {\it Athena} X-IFU and WFI observations of N132D and 1E 0102.2-7219 (E0102). The objective of these simulations is to determine if pileup is a significant issue for {\it Athena} observations of N132D or E0102 with  the X-IFU or WFI. A preliminary assessment suggests that pileup is not a concern but a more careful examination is warranted.  The group intends to work on the IACHEC N132D model
in the coming months so that a mature model may be released soon after the next IACHEC meeting in 2019.

The group continues to use the IACHEC model for E0102 (Plucinsky et al. 2017) to improve the calibrations of their respective instruments. S.\ Chandra used the E0102 model to identify issues with the gain correction on the \astrosat/SXT. He showed the SXT spectra with the revised gain correction that produces a dramatically improved fit at low energies.  P.\ Plucinsky showed the ACIS fits to E0102 from 2016 to 2018 with the contamination model (N0010) for \chandra/ACIS which indicated that the model was significantly overestimating the contamination layer for recent observations. The CXC has released two contamination models since the IACHEC meeting, one (N0011) to update the model for ACIS-I and another (N0012) to update the model for ACIS-S.  A.\ Beardmore showed the empirical model he has developed for Cas A (summed over the entire remnant) and how he uses it to characterize the charge traps and CTI in the \swift/XRT CCD.  M.\ Guainazzi commented that this empirical model for Cas A should be considered for inclusion in the set of IACHEC standard models.

\section*{References\footnote{see {\tt http://web.mit.edu/iachec/papers/index.html} for a complete list of IACHEC papers}}

\noindent
Bamba et al. 2018, ApJ, 854, 71\\
\noindent
Chen Y., et al., 2018, Journal of American Statistical Association, accepted, arXiv:1711.09429\\
\noindent
Hitomi Collaboration et al. 2018, PASJ, 70, 16\\
\noindent
Kaastra J. \& Bleeker J.A.M., 2016, A\&A, 587, 151 \\
\noindent
Kaastra, J.S., 2017, A\&A, 605, A51\\
\noindent
Madsen K., et al., 2017a, AJ, 153, 2 \\
\noindent
Marshall H., et al., 2004, SPIE, 5165 \\
\noindent
Nevalainen et al., 2010, A\&A, 523, A22\\
\noindent
O'Dell S., et al., 2013, SPIE, 8559 \\
\noindent
Plucinsky P., et al., 2016, Proc. SPIE, 9905, 44 \\
\noindent
Plucinsky P., et al., 2017, A\&A, 597, 35 \\
\noindent
Plucinsky P.,et al., 2018, Proc. SPIE, 10699\\
\noindent
Rasmussen, C.E.\ and Williams, C.K.I., 2006, {\sl Gaussian Processes for Machine Learning}, the MIT Press, ISBN 026218253X, {\tt http://www.gaussianprocess.org/gpml/chapters/}\\
\noindent
Terada Y., et al., 2008, PASJ, 60, 25\\

\end{document}